\documentclass[a4paper]{jpconf}
\usepackage{graphicx}
\newcommand\be{\begin{equation}}
\newcommand\eeq{\end{equation}}
\begin{document}
\title{Nuclear constraints on the equation of state and rotating neutron stars}

\author{A E L Dieperink$^1$ and C Korpa$^2$   }

\address{$^1$ Kernfysisch Versneller Instituut,
  NL-9747AA Groningen, The Netherlands}
\address{ $^2$ Department of Theoretical physics, University of P\'ecs, Ifj\'us\'ag \'utja 6, 7624 P\'ecs, Hungary}
\ead{dieperink@kvi.nl}

\begin{abstract}
In this contribution  nuclear constraints on the equation of state for a neutron star are discussed.  A combined fit to nuclear masses and charge radii leads to
improved values for the symmetry energy  and its derivative  at nuclear saturation density,
$S_{\rm{v}}= 31$~MeV and $L=68\pm 8$~MeV.
As an application  the sensitivity of some properties of rotating supramassive neutron stars on the EoS is discussed.
\end{abstract}

\section{Nuclear Constraints on the Equation of State}
\begin{figure}[h]
 \centering
\includegraphics[width=11cm]{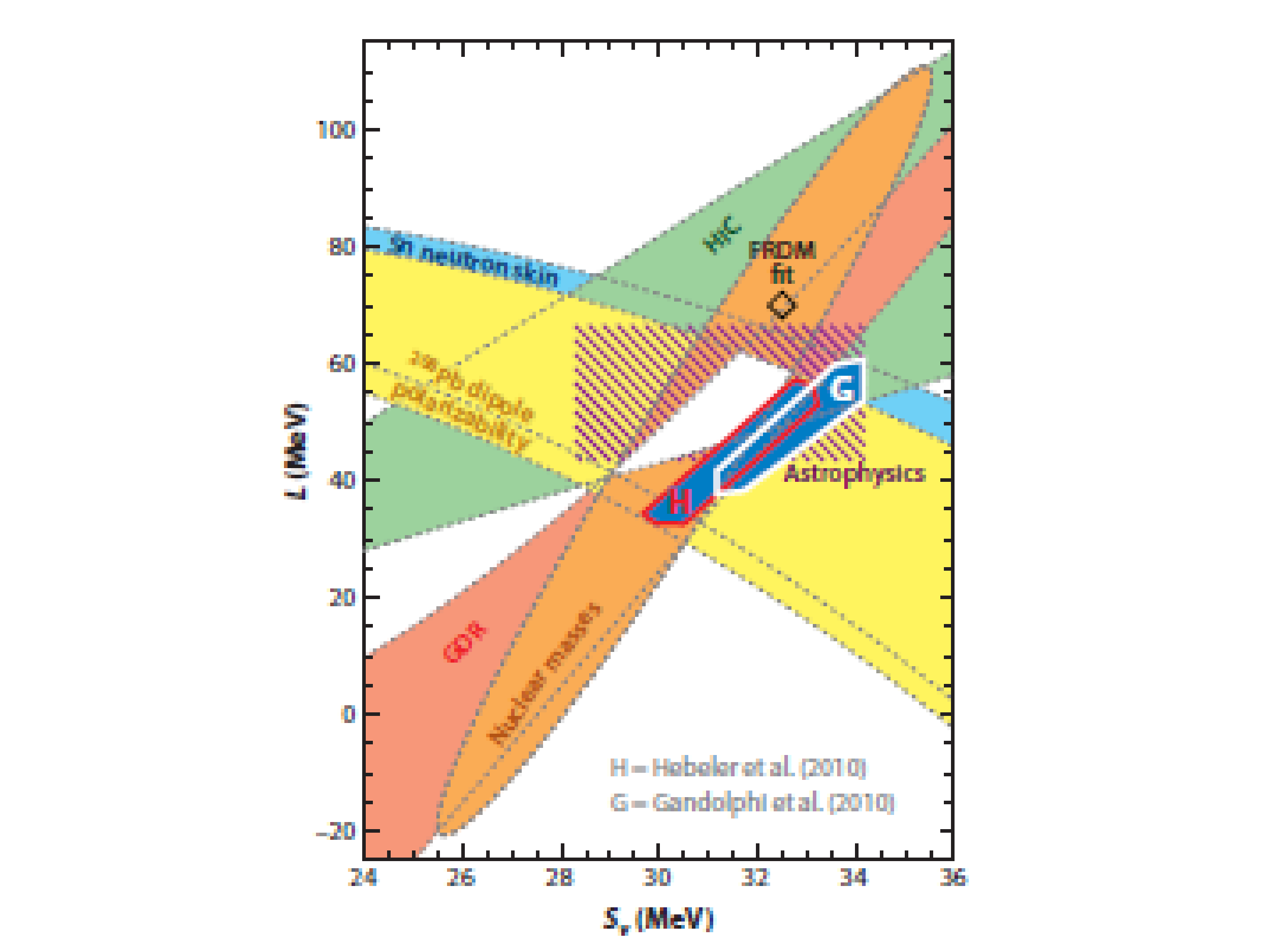}
\caption{Constraints on $L\ vs \ S_{\rm{v}}$, the filled ellipsoid corresponds to the constraint from fitting nuclear masses, from \cite{Lattimer12}.}
\label{f_SvsL}
\end{figure}

\begin{figure}
\includegraphics[width=7cm]{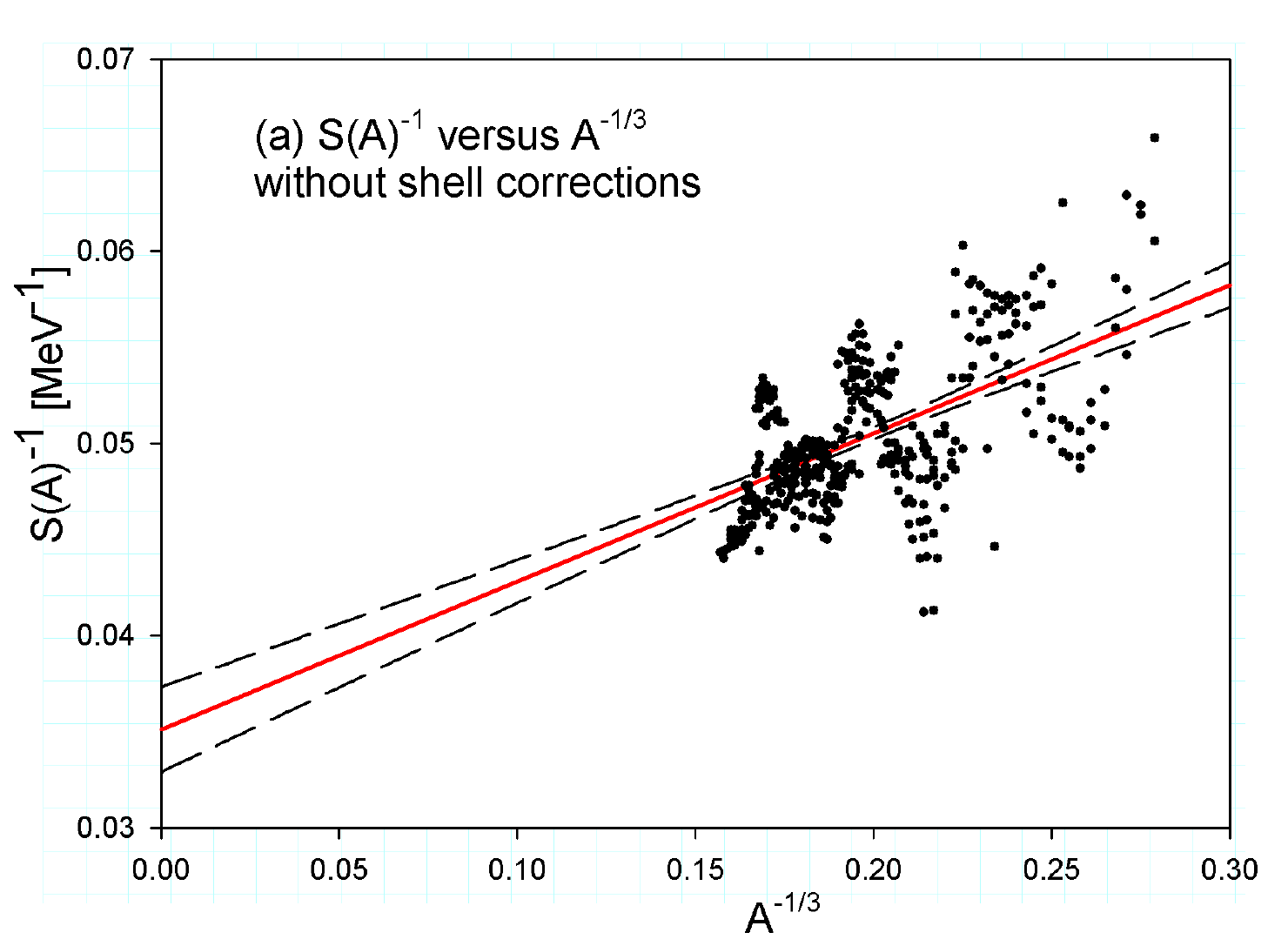}
\includegraphics[width=7cm]{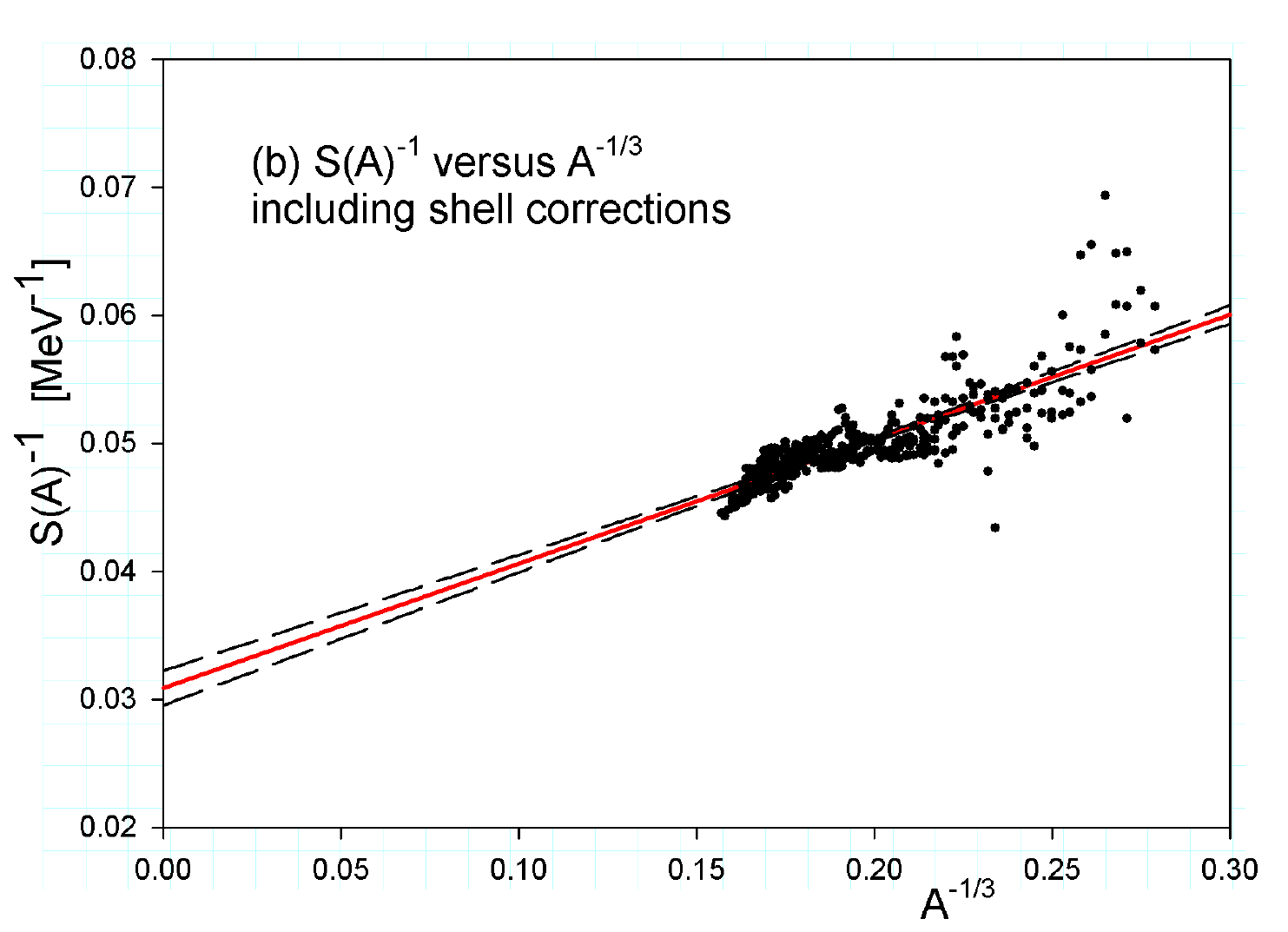}
\caption{SE vs $A^{-1/3}$ without (left) and with (right)  shell corrections, from \cite{Dieperink09}.}
\label{f_SEvsA}
\end{figure}

Despite numerous efforts to tighten the nuclear constraints on the EoS there remains considerable uncertainty.
The spreading in the pressure at nuclear saturation density as summarized by Lattimer~\cite{Lattimer12} a decade ago was roughly a factor six;
results of present day  mean field calculations  vary by about a factor four~\cite{Steiner12}.
Therefore it remains a challenge to try to improve the situation.
\\ The pressure as a function of density is given by $P(\rho)=\rho d\epsilon /d\rho -\epsilon(\rho).$
\\ In neighborhood of saturation density $\rho_s$, with  $u=\rho/\rho_s$ and $x$  proton fraction one has
\be \epsilon(u,x)=  B+  K/18(u-1)^2 + S_A(u)(1-2x)^2+.., \eeq
where $K$ the compressibility and $S_A$ the symmetry energy (SE). Hence the pressure near $\rho_s$ is
 \be P(\rho\sim\rho_{\rm{s}}) \sim u^2\rho_{\rm{s}} [\frac{K}{9}(u-1) + \frac{dS_A}{du}(1-2x)^2 +..]. \eeq
 In practice the leading contribution comes from the last term, the derivative of the SE;
the latter is usually parameterized in the liquid drop model (LDM) as
\be S_A= \frac{(N-Z)^2}{A} \frac{S_{\rm{v}}}{1+yA^{-1/3}}, \label{SE} \eeq
  where $S_{\rm{v}}, \ S_{\rm{s}}$ denote the volume and surface SE,   and $y=S_{\rm{v}}/S_{\rm{s}}$.
\\  The quantity of interest for the EoS, the derivative  $L= 3\rho_sdS/d\rho|_s$,  can in good approximation~\cite{Lattimer12} be related to $ S_{\rm{v}}$ and $y$:
  ${{y}} \sim 0.646+ S_{\rm{v}}/98$MeV$+0.436L/S_{\rm{v}}+0.087(L/S_{\rm{v}})^2 $.
\\ In practice the values of $S_{\rm{v}}, y$ when fitted to masses using the LDM appear to be strongly correlated \cite{Dan03},
   and the same is true for $L$, $S_{\rm{v}}$ (see the 1-$\sigma$ confidence ellips in fig.~\ref{f_SvsL}),
and a  similar correlation is found in  microscopic (mean field) models.
\\ However, one can improve the situation sketched above in several ways.
As a first step one may consider differentials of masses with respect to $N-Z$ (rather than a global fit),  which allows one to fit  the parameters in $\frac{S_{\rm{v}}}{1+yA^{-1/3}} $ in isolation of other terms \cite{Dieperink09,Dan09}.
\\ By plotting   $1/S$ vs $A^{-1/3}$  (see fig.~\ref{f_SEvsA}) one obtains the value of $1/S_{\rm{v}}$
from the crossing of the fit line with y-axis ($A=\infty$) and the slope $y= 2.6\pm 0.8$.
 From   the figure the  correlation between slope and $S_{\rm{v}}$ is evident.
\\ As a second step  an appreciable increase in the accuracy can be achieved by including shell corrections~\cite {Dieperink09}  shown in the right part of the figure.
 In passing we note that in ref.~\cite{Jiang12} a quite accurate result for  $S_{\rm{v}},\ S_{\rm{s}}$ is reported by using double differences of masses. However, their results are obtained for a  parametrization of the SE different from eq.~(\ref{SE}), and moreover depend on the choice of the Wigner energy.
\\ As a final step one can improve the situation further by using information from charge radii, which mainly depends only on the ratio $y$.
(In fig.~\ref{f_SvsL} this is indicated by the band labeled ``skins of Sn", but we consider the result of this particular analysis rather model dependent).
\\ In the spirit of the LDM and  distinguishing proton and neutron radii we decompose \cite{Dieperink09}
\be R_{\rm{p}}(N,Z)=R_0(N,Z) +\frac{N}{A} R_{\rm{np}}(N,Z)+ \delta R_{\rm C}(N,Z), \label{Rcharge} \eeq
 where the isoscalar term $R_0= (NR_{\rm n}+ZR_{\rm p})/A \sim r_0A^{1/3}, $ and the
  isovector term (essentially the neutron skin)
\be R_{\rm{np}}=R_{\rm n}-R_{\rm p} =\frac{2r_0}{3}\frac{N-Z}{A}\frac{1}{1+  A^{1/3}/y}, \eeq
and $R_C$ the Coulomb contribution~\cite{Dan03}.
 The  point is that $R_{\rm{np}}$ depends only on $y=S_{\rm{v}}/S_{\rm{s}}$ (apart from the Coulomb contribution).
\\ To determine $y$ from data one can envision the following options
\\   (i) measure the neutron skin using parity violating electron scattering (PREX). However, the first experiment \cite{PREX} on $^{208}$Pb yielded a
 rather large error  $R_{\rm{np}} \sim 0.33 \pm 0.17$ fm.
  \\  (as a side remark: atomic parity violation, in progress, appears a promising alternative tool, with a
  possible precision of about 1\% in the skin in Ra isotopes),
\\ (ii) fit to observed  charge radii using the  expression (\ref{Rcharge}).
(As an alternative one may  consider fitting differences like isobar shifts,
 $R_{\rm p}(N,Z)-R_{\rm p}(N-i,Z+i)$; the latter are independent of $R_0$, but in general  have larger experimental uncertainties.)
\\ The values for $S_{\rm{v}}$ and $y$ from a combined fit of masses and radii are given in table~\ref{table1}, which are compared to some other results from  fits and microscopic approaches.
\begin{table}[!h]
\caption{Results for the parameters $S_{\rm{v}},\ y$ and $ L$ obtained from fits and microscopic approaches}
\begin{tabular}{llllll}
\br
 $S_{\rm{v}}$(MeV)  & $y$  & $L$(MeV)   &  $R_{\rm{np}}$($^{208}$Pb) (fm)  &             ref & model \\
 \mr
        &            &  {{fit to masses}}   &                &      &    \\
 \mr
$32.5\pm 0.5$ &   1.98  & $70\pm 15$   &             &  \cite{Moller12}    & FRLDM\\
$32.1\pm 0.3$ & 1.9     &              &             &  \cite{Jiang12} &  double diff\\
31.1 & $2.31\pm 0.38$    & $66\pm 13$   &             &     \cite{Liu}  &  LDM    \\
 32  &  3.0    &   94       &             &  \cite{Dan09} &  analysis IAS\\
 31  &  $2.5\pm 0.4$ & $ 80\pm 15$ &         &    \cite{Dieperink09}& LDM+shell corr\\
 30.5 & $ 2.35 \pm 0.20$ & $68\pm 8$  &  $0.185\pm 0.015$ & present & masses+ charge radii
\\ \mr
   &                   &  Microscopic & approaches          & & \\
   \mr
30           &         &   $58\pm 15$       &         &   \cite{Chen12} & Skyrme+skin Sn isotopes \\
$32\pm 1.5$  &         &        $45\pm 15$     &        &    \cite{Gandolfi}     & QMC \\
$31\pm 1 $   & $1.85\pm 0.25$  &  $46\pm10$  &             &  \cite{Hebeler}  &  EFT \\
 $ 31 \pm 1$  &        & $ 64\pm 8$        &  $0.195\pm 0.02$ &   \cite{Agrawal}  & EDF
\\ \br
 \end{tabular}

 \label{table1}
 \end{table}
 Note that phenomenology seems to favor larger $L$ values than most microscopic models; this trend is not understood yet.

\section{Rotating supramassive neutron stars and the EoS}
\begin{figure}[!b]
 \centering
\includegraphics[width=9cm]{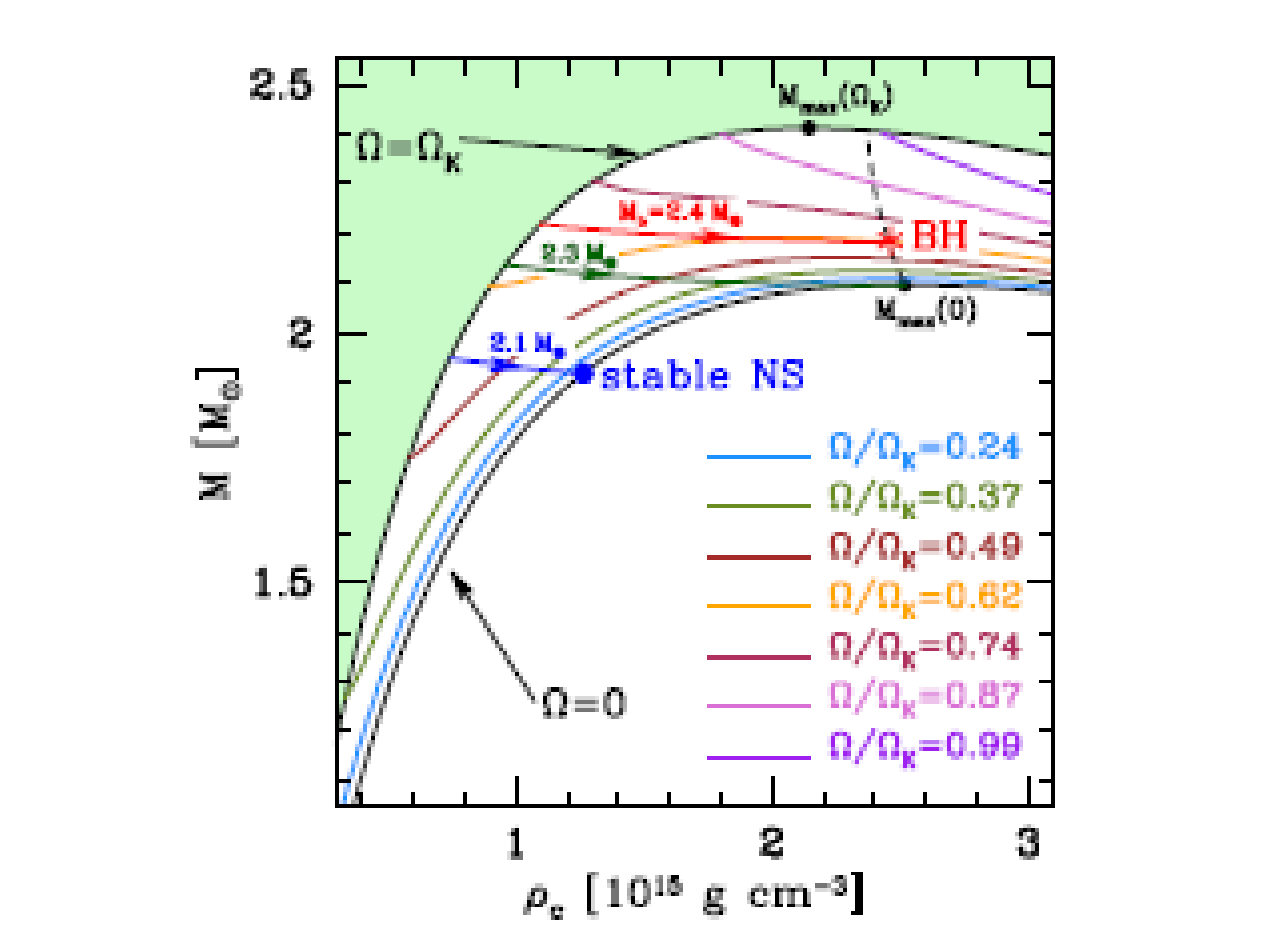}
\begin{minipage}[b]{6cm}
\caption{Gravitational mass {\it vs} central density for various ratios $f=\Omega/\Omega_K$, where $\Omega_K$ is taken as the critical $\Omega$ belonging to star with maximum mass; arrows represent tracks of NS's slowing down due to magnetic braking, from {\cite{Falcke13}}.}
\end{minipage}
\label{rotns-FR}
\end{figure}

 Several energetic observations can be associated with formation  of neutron stars (NS) or black holes (BH),
  supernovae, gamma ray bursts (GRB).
 Some short GRB's  ($\sim$1s) have been attributed to NS mergers.
\\ Recently the observation \cite{Thornton13} of bright radio pulses was reported, with radio flux $\sim $ Jy at GHz frequencies and $\Delta T \le 1 $ms, which
 do not repeat, while no $\gamma$- or x-rays were observed.
  \\   Falcke and Rezzolla \cite{Falcke13} proposed the following interpretation: a  supramassive rotating NS (i.e., a NS with a mass larger than the maximum mass of a static NS,
  {\it{e.g.}} created by accretion in binary system) slows down due to magnetic braking,
 and at critical point collapses into Kerr BH (see fig.~\ref{rotns-FR}).
The created event horizon will hide star's surface, hence only emission from the detached magnetosphere can be observed;
 the estimated timescale (freefall) $\tau \sim 0.04  R_{10}^{3/2}M_2^{-1/2}$ ms appears to be consistent with the observation.
\\ For a non-rotating star the mass vs radius relation (given the EoS) is obtained by solving the TOV equation; a rotating star requires a more general approach to general relativity.
 \begin{figure}[!h]
\includegraphics[width=7.5cm]{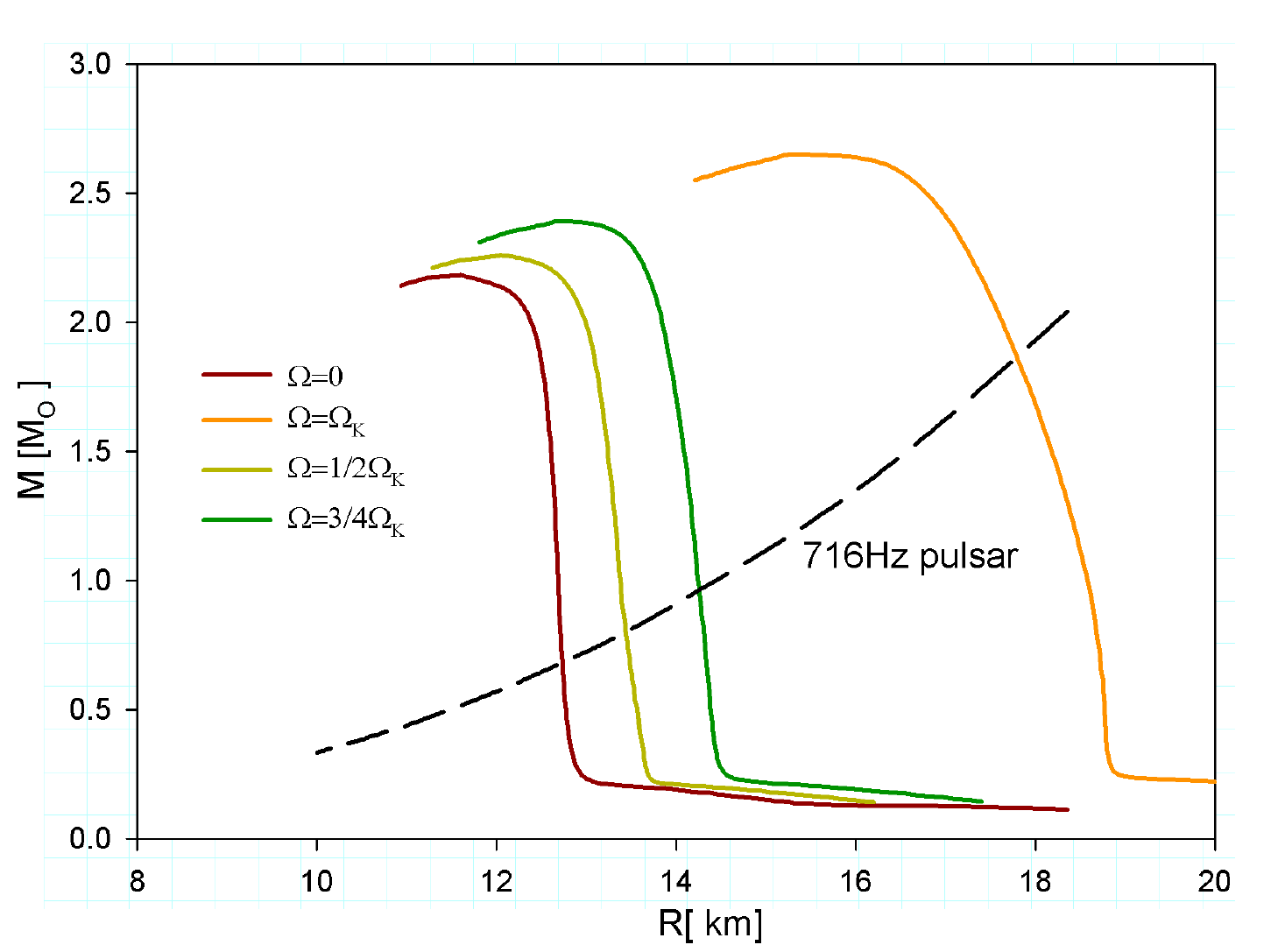}
\includegraphics[width=7.7cm]{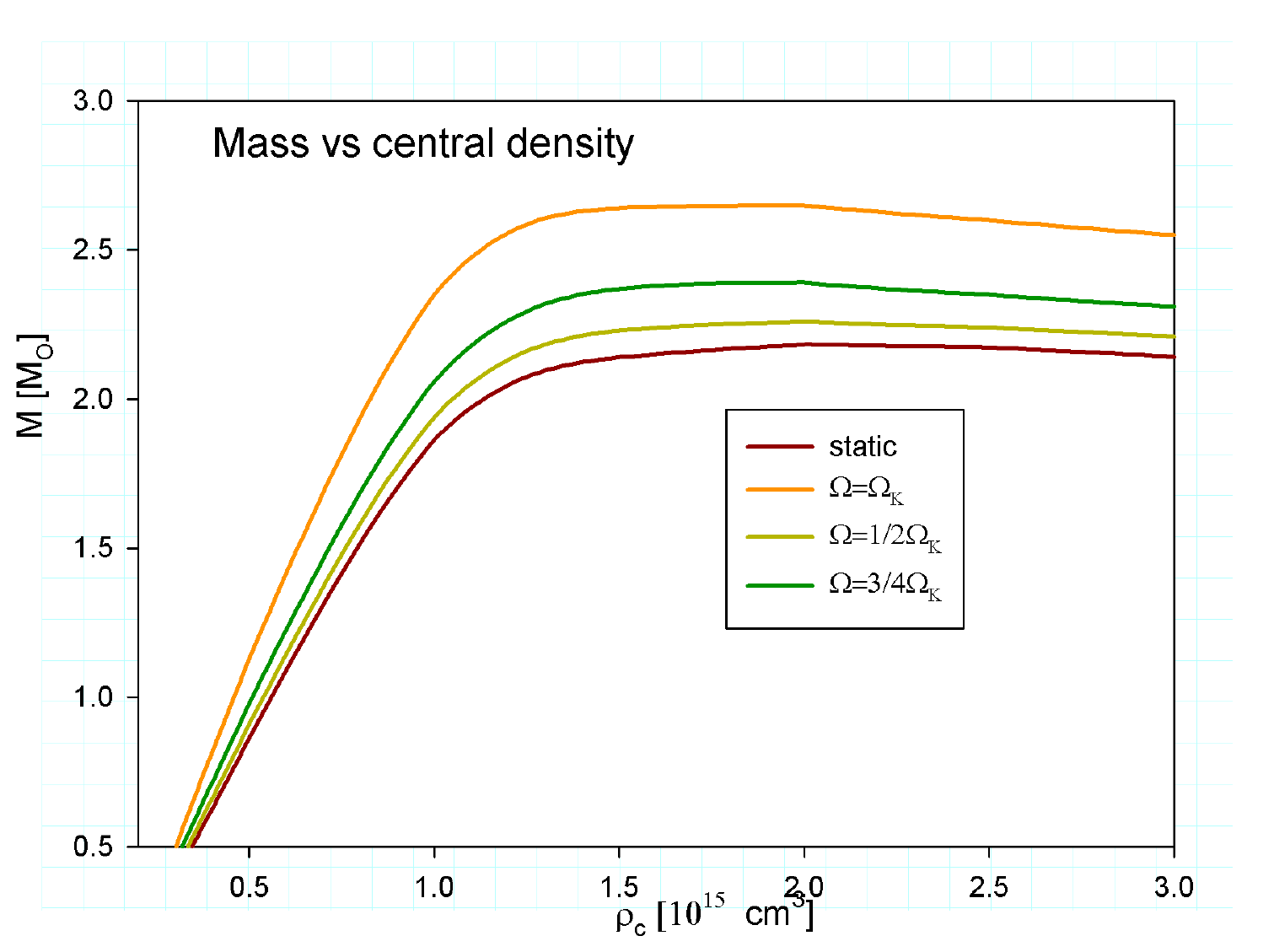}
\caption {Gravitional mass {\it{vs}} radius (left) and {\it{vs}} central density (right) for various $\Omega$.
In the left panel the broken line denotes the lower limit on the mass of a static star from eq.~(\ref{taumin}). }
\label{nsrot}
\end{figure}
The scenario proposed in ref.~\cite{Falcke13} seems not unrealistic but some questions remain.
For example the authors took a very simple representation for the
  EoS: a single polytrope  $ P={{K}}\rho^\gamma$ with  $\gamma=2$ and  $K$ adjusted such that  $M(0)_{\rm{max}} =2.1M_\odot$.
\\ Naturally one may ask how large is the sensitivity to EoS?
 To investigate this
we took a 3-polytrope EoS , $P(\epsilon)=K_i\epsilon^{\Gamma_i}$. It fits  $P(\rho_s)$ and has the proper low-density behavior; specifically,  $\Gamma_1=1.5$ for  $\epsilon < 67 $
MeV/fm$^3$, $\Gamma_2=2.68,$ for $67 < \epsilon < 650 $, and  $\Gamma_3= 1.41$ for $\epsilon > 650 $.
The $K_i$ values are fixed by continuity of the pressure $P$ and the normalization $P(\epsilon=650)=180$ MeV/fm$^3$. It yields    $M_{\rm {max}}=2.2M_\odot$.
\\ Using the rns code \cite{Stergioulas95}  the mass as a function of the equatorial radius or the central density, and the critical frequency (the Keppler or mass shedding limit) have been computed, see fig.~\ref{nsrot}.
\\ Qualitatively the main features of fig.~\ref{rotns-FR} are confirmed,  i.e., $M(\Omega_K)$ increases by 20$\%,$
 \   $R(\Omega_K)$ increases by 50\%.
 However, $\Omega_K$ itself turns out to be more sensitive to the EoS \cite{Haensel,Lo11}.
 \\ Finally we point out that an observation of a high rotation frequency of a pulsar can lead to a constraint on the mass-radius diagram. Namely for a  Newtonian uniformly rotating rigid star with mass $M$ and radius $R$
one has  \be \tau_{\rm{min}}= 2\pi\sqrt{\frac{R^3}{GM}} =0.545 \left(\frac{M_\odot}{M}\right)^{1/2} \left(\frac{R}{\rm{10km}}\right)^{3/2} \rm{ms}. \eeq
  Using general relativity
 a similar empirical relation (valid for $M<0.9 M_{\rm{max}}(0)$   with a weak dependence on the EoS) has been derived \cite{Haensel} ($M,  R$  refer to the static star)
\be \tau_{\rm{min}} = C \ (M_\odot/M)^{1/2} (R/{\rm{10km}})^{3/2} \rm{ms} \label{taumin},
 \eeq
 where $C\sim 0.92\pm 0.04$; with our EoS we find $C=0.90$.
 At present, the fastest rotating pulsar has $\nu_{obs}$ = 716 Hz. Obviously, the Keplerian
frequency of any neutron star must satisfy  $\nu_K \geq \nu_{\rm{obs}}.$ This inequality constraints a region on the $M \ vs\  R$ diagram as shown in the left panel of fig.~\ref{nsrot}.

\end{document}